# BROAD ABSORPTION LINE VARIABILITY IN REPEAT QUASAR OBSERVATIONS FROM THE SLOAN DIGITAL SKY SURVEY

Britt F. Lundgren,[1] Brian C. Wilhite,[1,2] Robert J. Brunner,[1,2] Patrick B. Hall,[3] Donald P. Schneider,[4] Donald G. York,[5,6] Daniel E. Vanden Berk,[4] and Jonathan Brinkmann[7]

## ABSTRACT

We present a time-variability analysis of 29 broad absorption line quasars (BALQSOs) observed in two epochs by the Sloan Digital Sky Survey. These spectra are selected from a larger sample of BALQSOs with multiple observations by virtue of exhibiting a broad C IV $\lambda1549$ absorption trough separated from the rest frame of the associated emission peak by more than 3600 km s$^{-1}$. Detached troughs facilitate higher precision variability measurements, since the measurement of the absorption in these objects is not complicated by variation in the emission-line flux. We have undertaken a statistical analysis of these detached-trough BALQSO spectra to explore the relationships between BAL features that are seen to vary and the dynamics of emission from the quasar central engine. We have measured variability within our sample, which includes three strongly variable BALs. We have also verified that the statistical behavior of the overall sample agrees with current model predictions and previous studies of BAL variability. Specifically, we observe that the strongest BAL variability occurs among the smallest equivalent width features and at velocities exceeding 12,000 km s$^{-1}$, as predicted by recent disk-wind modeling.

*Subject heading:* quasars: general

*Online material:* extended figure set

## 1. INTRODUCTION

Broad absorption lines (BALs) in quasar spectra are thought to be caused by obscuring overdensities in a wind, which is likely produced near the central accretion disk and propagated radially at some angle above an optically thick torus (e.g., Murray et al. 1995; Elvis 2000; Proga et al. 2000). BALs, which by definition exhibit a full width at half-maximum (FHWM) of more than 2000 km s$^{-1}$, have been observed at velocities as great as 66,000 km s$^{-1}$ in the quasar rest frame (Foltz et al. 1983). Mini-BALs compose a similar class of somewhat narrower lines, also considered intrinsic to the quasar, which are characterized by a FWHM in the range of 300–2000 km s$^{-1}$ (Hamann & Sabra 2004). In general, BALs are observed in 15%–20% of quasars (Weymann et al. 1991; Reichard et al. 2003). This percentage is typically interpreted as the fraction of the viewing angle obscured by these overdensities, although the occurrence of BALs may instead indicate a particular phase of quasar evolution (Green et al. 2001; Yuan & Wills 2003).

The accretion process that fuels the central engine of a quasar is prone to dramatic activity, presumably due to both internal turbulence in the surrounding accretion disk and the strong gravitational effects of a supermassive black hole. As a result, stochastic variations in the radiative output of quasars have been observed (e.g., Vanden Berk et al. 2004; Wilhite et al. 2005), and such variability can be expected to propagate into the observational properties of any matter confined within the disk wind (Murray et al. 1995; Elvis 2000; Proga et al. 2000). Variability of the equivalent widths of intrinsic absorption lines such as BALs may indicate a changing level of ionization in regions very near to the quasar central engine or a change in the covering factor or optical depth of the absorbing gas. Changes in the velocity separation between absorption features and corresponding emission lines can provide valuable insight into the wind acceleration mechanism, which in turn can help to illuminate the overall structure and physical properties of the quasar.

Data from large surveys such as the Sloan Digital Sky Survey (SDSS; York et al. 2000) have significantly increased the number of known BALQSOs in the past decade. While large catalogs of BALQSOs now exist (Reichard et al. 2003; Trump et al. 2006), to date, minimal work has been done with regard to their variability. Narrow intrinsic absorption line variability has already been detected in a number of studies (e.g., Hamann et al. 1995, 1997; Ganguly et al. 2001; Wise et al. 2004; Misawa et al. 2005), and BAL variability has been previously observed in a few objects (Foltz et al. 1987; Turnshek et al. 1988; Smith & Penston 1988; Barlow et al. 1992; Vilkoviskij & Irwin 2001; Ma 2002). Some studies have reported equivalent width variability in ~25%–33% of these broad features (Barlow 1994; Narayanan et al. 2004), yet no large statistical variability study has been published. The time-variability analysis by Barlow (1994), which examines 23 BALQSOs, is the largest study of this kind to date. Repeat spectroscopic observations from the SDSS provide a unique opportunity to study the variation in these sources, since the sample size is large and the homogeneity of SDSS data allows for precise measurements of line variability.

In this paper, we analyze a sample of 29 SDSS quasars with two-epoch observations and prominent C IV $\lambda1549$ BALs that are effectively detached from the associated emission peak. Although our sample is considerably smaller than the latest BAL quasar samples, this work is the largest comprehensive spectroscopic variability study of BALQSOs to date.

[1] Department of Astronomy, University of Illinois at Urbana-Champaign, Urbana, IL.
[2] National Center for Supercomputing Applications, Champaign, IL.
[3] Department of Physics and Astronomy, York University, Toronto, ON, Canada.
[4] Department of Astronomy and Astrophysics, Pennsylvania State University, University Park, PA.
[5] Department of Astronomy and Astrophysics, University of Chicago, Chicago, IL.
[6] Enrico Fermi Institute, University of Chicago, Chicago, IL.
[7] Apache Point Observatory, Sunspot, NM.

## 2. OBSERVATIONS AND DATA REDUCTION

### 2.1. The Sloan Digital Sky Survey

Through 2005 June, the SDSS had imaged 8000 deg$^2$ and obtained follow-up spectra for nearly $7 \times 10^5$ galaxies and $9 \times 10^4$ quasars. Imaging data are acquired by a 54-chip drift-scan camera (Gunn et al. 1998, 2006) on the dedicated 2.5 m telescope at Apache Point Observatory in New Mexico. The data are reduced and calibrated by the PHOTO software pipeline (Lupton et al. 2001). The photometric system is normalized such that the SDSS $u$, $g$, $r$, $i$, and $z$ magnitudes (Fukugita et al. 1996) are on the AB system (Smith et al. 2002). A 0.5 m telescope monitors site photometric quality and extinction (Hogg et al. 2001). Point-source astrometry for the survey is accurate to within less than 100 mas (Pier et al. 2003), and imaging quality control is discussed in Ivezić et al. (2004).

A fraction of the objects located in the imaging are targeted for follow-up spectroscopy as candidate galaxies (Strauss et al. 2002; Eisenstein et al. 2001), quasars (Richards et al. 2002), or stars (Stoughton et al. 2002). Targeted objects are grouped in 3° diameter tiles (Blanton et al. 2003), and aluminum plates are drilled with 640 holes at positions corresponding to the objects' sky locations. When the telescope is in spectroscopic mode, plates are placed in the imaging plane of the telescope and plugged with optical fibers that run from the telescope to twin spectrographs. Roughly 500 galaxies, 50 quasars, and 50 stars are observed on each plate, and the remaining fibers are utilized for sky subtraction and calibration.

SDSS spectra cover the observer-frame optical and near-infrared, from 3900 to 9100 Å, with a resolution of $\lambda/(\Delta\lambda) \approx 2000$ at 5000 Å (Stoughton et al. 2002). Spectra are obtained in three or four consecutive 15 minute observations until an average minimum signal-to-noise ratio is met. Observations of 32 sky fibers, eight reddening standard stars, and eight spectrophotometric standard stars are used to calibrate the science targets' spectra. The Spectro2d pipeline flat-fields and flux calibrates the spectra, and the Spectro1d code identifies spectral features and classifies objects by spectral type (Stoughton et al. 2002). Ninety-four percent of all SDSS quasars are identified spectroscopically by this automated calibration; the remaining quasars are identified through manual inspection. Quasars are defined to be those extragalactic objects with broad emissions lines (FWHM velocity width $\gtrsim 1000$ km s$^{-1}$ regardless of luminosity).

For this paper, we use spectra with additional calibration, as described in Wilhite et al. (2005). Through 2004 June, objects corresponding to 181 plates had been observed multiple times, with time lags between observations ranging from days to years. These second observations, although originally unplanned, came as the product of normal survey operations, usually early in the survey, when a method had not yet been developed for determining when a plate had reached an adequate signal-to-noise ratio. As discussed in Wilhite et al. (2005), spectra from plates observed greater than 50 days apart are guaranteed not to have been co-added and are more suitable for use in variability studies. There are 53 such large time-lag plate pairs containing almost 2200 quasars; single-epoch observations of 52 of these plate pairs are contained in the Fourth Data Release (DR4; Adelman-McCarthy et al. 2006).

### 2.2. Refinement of Spectroscopic Calibration

As was shown in Vanden Berk et al. (2004) and Wilhite et al. (2005) some additional spectrophotometric calibration of SDSS spectra is necessary for variability studies. The calibration methods used are summarized below; see Wilhite et al. (2005) for a complete discussion. The Spectro1d pipeline calculates three signal-to-noise ratios (S/Ns) for each spectrum by calculating the median S/N per pixel in the regions of the spectrum corresponding to the SDSS $g$, $r$, and $i$ filter transmission curves. Hereafter, when referring to the two halves of a plate pair we use the phrase "high-S/N epoch" to refer to the plate with the higher median $r$-band S/N. The plate with the lower median $r$-band S/N is called the "low-S/N epoch." This is a plate-wide designation; although most objects follow the plate-wide trend, this does not speak to the relative S/N values for any given individual object, nor does it correspond to an object's relative line or continuum flux at a given epoch. The stars on a plate are used to resolve calibration differences between the high- and low-S/N epochs, under the assumption that the majority of stars are nonvariable (precautions are taken to remove the obviously variable stars from recalibration; see Wilhite et al. [2005] for complete details). For each plate pair, we create a recalibration spectrum, equal to the ratio of the median stellar high-S/N epoch flux to the median stellar low-S/N flux, as a function of wavelength. This recalibration spectrum is fitted with a fifth-order polynomial to preserve real wavelength dependences but remove pixel-to-pixel noise (cf. Fig. 5 of Wilhite et al. 2005), leaving a smooth, relatively featureless curve as a function of wavelength. All low-S/N epoch spectra are then scaled by this "correction" spectrum.

### 2.3. BALQSO Sample Selection

Each observation at both epochs of the 2200 objects compiled by Wilhite et al. (2005) was inspected by eye for the presence of C IV BAL and mini-BAL features. BALQSO spectra were selected for further study if they contained at least one C IV absorption feature with a velocity width of at least 1000 km s$^{-1}$ in one or more epochs with no discernible doublet structure. Such doublets might indicate that the absorption is due to a collection of overlapping narrow C IV absorption lines rather than one broad absorber (Weymann et al. 1991).

Objects meeting these primary criteria were also tested to ensure that intervening absorption systems were not responsible for any of the apparent BALs, since Mg II $\lambda 2800$ and Fe II $\lambda\lambda 2261$, 2344, 2374, 2383, 2587, 2600 in multiple lower redshift absorption line systems have the potential to overlap and produce seemingly broad features in the region blueward of C IV. Such occurrences are easily recognized, as the Mg II doublet exhibits a significantly larger rest-frame wavelength separation (2796–2803 Å) than that of C IV (1548–1551 Å). Furthermore, Fe II absorption lines occur over a large wavelength range in the visible spectrum and are almost always accompanied by a strong Mg II doublet with the same redshift. These characteristics ensure that even if low-ionization absorption lines are present in the wavelength range where intrinsic C IV absorption would be observed, these intervening absorbers may be accurately identified and removed from the C IV BALQSO sample.

Of the 1186 objects in this sample within the redshift range where the C IV emission line can be measured in the SDSS spectra ($1.6 \leq z_{qso} \leq 4.2$), 172 (14.5%) contain BALs that met the aforementioned criteria. This percentage of BALs within the quasar population agrees with the expected values from previous catalogs (see, e.g., Reichard et al. 2003; Hewett & Foltz 2003).

Measurements of BAL variability are difficult to make, since in most cases the intrinsic nature of these lines produces an overlap with the quasar emission lines. This complication demands precise continuum and line profile fitting if one wishes to study the variability of absorption in the region of broad emission. It has also been established that the continua of quasars are prone to much greater variability than the emission-line flux (Wilhite et al. 2005).

Therefore, any efforts to study the relationship between quasar source variability and changes in the intrinsic absorption features must cleanly separate the effects of variability in the quasar continuum and emission lines. By restricting our sample to those BALs that do not overlap with the broad C IV emission feature, we can obtain higher precision measurements of the equivalent widths of these features and directly probe the correlation between quasar continuum emission and BAL variability.

As a result, we narrowed our initial sample of 172 multi-epoch SDSS BALQSOs to 29 that were found to contain broad absorption lines separated from the associated C IV emission line by at least 3600 km s$^{-1}$ in the quasar rest frame. Vanden Berk et al. (2001) finds 1546.15 Å to be the median wavelength of C IV emission in the composite spectrum of 2200 SDSS quasars and a line width ($\sigma_\lambda$) of 14.33 Å. On the blue side of the emission peak, this width is approximately equivalent to a velocity of 3600 km s$^{-1}$ with respect to the C IV laboratory rest-frame wavelength of 1549 Å. This lower limit separation ensures that the absorption feature lies on a locally flat section of the spectrum that, at most, only marginally overlaps with the C IV emission region. Within these 29 spectra, we find 33 detached-trough BALs meeting the 3600 km s$^{-1}$ criterion, as some quasars host more than one distinct BAL at high velocity. The observational details of the selected quasars are presented in Table 1.

## 3. DATA ANALYSIS

Although there has been some success in automating the identification of BALs from within a large sample of quasar spectra (Trump et al. 2006), the actual analysis of the BAL features within these spectra requires manual inspection. The inability to fully automate this process is largely due to the tendency of BAL widths and velocities in the quasar rest frame to vary greatly between objects. Furthermore, the continuum fitting, on which any absorption profiles must be based, is complicated by the very presence of the BALs blueward of the emission peaks. For these reasons, each spectrum requires careful analysis to ensure that the wavelength ranges included in the continuum fitting and equivalent width measurements of the BAL features are appropriate. The rest of this section details the processes that we used to perform measurements of the BAL features.

### 3.1. Continuum Fitting

To determine the appropriate continuum level we first deredshifted each spectrum into the rest frame of the quasar. The SDSS pipeline occasionally calculates a different redshift for the two epochs (e.g., due to S/N differences or incorrect emission-line identifications). For consistency, we always use the SDSS redshift assigned to the higher S/N spectrum. Wilhite et al. (2006) showed that the shift in central wavelength of the C IV emission peak between epochs is negligible, even in variable quasars. Thus, a single high-S/N calculation of the emission redshift should be satisfactory for both epochs. Since our goal is the measurement of variability in the absorption features, it is imperative that the redshift of the emitting source be consistent between observations, which ensures that all measurements incorporating velocity use the same standard scaling in both epochs.

After deredshifting each spectrum into the rest frame of the quasar, a continuum is fit across the region of C IV emission. Because the part of the spectrum analyzed here is relatively small (roughly 1400–1700 Å in the rest frame of the quasar), the continuum is approximated by a linear fit in wavelength in the quasar's rest frame, as was shown to be sufficient by Wilhite et al. (2006).

TABLE 1
SDSS Spectral Observations

| Object[a] | SDSS J | $z_{hsn}$[b] | MJD$_{hsn}$ | MJD$_{lsn}$ |
|---|---|---|---|---|
| 1................ | 115031.03−004403.1 | 2.3921 | 51,943 | 51,662 |
| 2................ | 130058.13+010551.5 | 1.9015 | 51,689 | 51,994 |
| 3................ | 132304.58−003856.5 | 1.8267 | 51,984 | 51,665 |
| 4................ | 131853.45+002211.5 | 2.0737 | 51,984 | 51,665 |
| 5................ | 132742.92+003532.6 | 1.8736 | 51,959 | 51,663 |
| 6................ | 133150.51+004518.7 | 1.8869 | 51,955 | 51,662 |
| 7................ | 134544.54+002810.7 | 2.4528 | 51,943 | 51,666 |
| 8................ | 143641.24+001558.9 | 1.8659 | 51,637 | 51,690 |
| 9................ | 144959.96+003225.3 | 1.7217 | 51,994 | 51,666 |
| 10.............. | 150206.66−003606.9 | 2.1998 | 51,990 | 51,614 |
| 11.............. | 150109.13−011502.7 | 2.1278 | 51,990 | 51,614 |
| 12.............. | 170633.06+615715.0 | 2.0115 | 51,695 | 51,780 |
| 13.............. | 172001.31+621245.7 | 1.7642 | 51,694 | 51,789 |
| 14.............. | 234506.31+010135.5 | 1.7969 | 51,877 | 51,783 |
| 15.............. | 031828.90−001523.1 | 1.9820 | 51,929 | 51,821 |
| 16.............. | 004527.68+143816.1 | 1.9842 | 51,868 | 51,812 |
| 17.............. | 093620.52+004649.2 | 1.7175 | 52,314 | 52,027 |
| 18.............. | 131305.74+015926.9 | 2.0169 | 52,295 | 52,029 |
| 19.............. | 081822.63+434633.8 | 2.0430 | 52,207 | 51,959 |
| 20.............. | 081416.75+435405.0 | 1.7036 | 52,207 | 51,959 |
| 21.............. | 153703.94+533219.9 | 2.4035 | 52,374 | 52,442 |
| 22.............. | 224019.01+144435.5 | 2.2424 | 52,520 | 52,264 |
| 23.............. | 143130.04+570139.0 | 1.8006 | 52,346 | 52,433 |
| 24.............. | 144403.97+565751.3 | 1.8544 | 52,347 | 52,435 |
| 25.............. | 145428.52+571441.3 | 3.2583 | 52,347 | 52,435 |
| 26.............. | 160649.24+451051.6 | 2.8256 | 52,443 | 52,355 |
| 27.............. | 102250.16+483631.1 | 2.0658 | 52,347 | 52,674 |
| 28.............. | 075010.17+304032.3 | 1.8954 | 52,346 | 52,663 |
| 29.............. | 081657.55+060441.7 | 2.0111 | 52,962 | 52,737 |

Note.—Subscripts "hsn" and "lsn" refer to high- and low-S/N epoch observations, respectively.

[a] These object numbers are used for internal reference and correspond to the labeling of individual quasars in Tables 2 and 3 and Fig. Set 7.

[b] The quasar redshift from the SDSS for the high-S/N epoch observation.

For each object, two regions of the spectrum are included to construct this fit, one on either side of the C IV emission line.

In most cases, the part of the spectrum corresponding to the quasar rest-frame wavelength range 1440–1460 Å was used to fit the region blueward of intrinsic C IV features. However, in cases in which this wavelength range contains a BAL, we choose a region at a longer wavelength that was still blueward of the C IV emission. As the location of the fit regions can vary from object to object, we present the wavelength boundaries used for the blue side of each fit in Table 2.

Significant emission from other ions, such as Fe II, is often found just redward of the C IV emission line (e.g., Wilkes 1984; Boyle 1990; Laor et al. 1994; Vanden Berk et al. 2001). As a result, the red side of the fit must be extended to longer wavelengths, where the spectrum returns to near-continuum levels (for a detailed discussion, see Wilhite et al. 2006). We have used the rest-frame region 1680–1700 Å to fit the red side of the continuum, as was done by Wilhite et al. (2006). Since 1680–1700 Å was always taken to be the range for the red component of these linear fits, this information has not been presented with the other fit parameters in Table 2.

Since BALs are thought to arise from absorption of gas at high velocities with respect to the emission-line source, each spectrum (and each corresponding wavelength region to be fit) is translated into velocity space zeroed to the central C IV emission

TABLE 2
Continuum Fitting Parameters (in Quasar Rest Frame)

| Object[a] (1) | $\lambda_{b1}$ (Å) (2) | $\lambda_{b2}$ (Å) (3) | $m_1$ $(10^{-5})$ (4) | $m_2$ $(10^{-5})$ (5) | $f_c(0)_1$ (6) | $f_c(0)_2$ (7) |
|---|---|---|---|---|---|---|
| 1 | 1440 | 1460 | $-0.95 \pm 0.31$ | $-0.33 \pm 0.22$ | $1.27 \pm 0.07$ | $1.84 \pm 0.05$ |
| 2 | 1430 | 1450 | $-0.39 \pm 0.31$ | $-0.25 \pm 0.38$ | $6.68 \pm 0.08$ | $6.45 \pm 0.10$ |
| 3 | 1440 | 1460 | $-4.40 \pm 0.80$ | $-3.55 \pm 0.52$ | $11.42 \pm 0.21$ | $11.93 \pm 0.13$ |
| 4 | 1430 | 1445 | $-0.50 \pm 0.39$ | $-0.82 \pm 0.32$ | $9.11 \pm 0.10$ | $8.00 \pm 0.08$ |
| 5 | 1440 | 1460 | $-0.86 \pm 0.53$ | $-2.12 \pm 0.41$ | $7.35 \pm 0.13$ | $8.00 \pm 0.10$ |
| 6 | 1440 | 1460 | $-1.87 \pm 0.43$ | $-1.57 \pm 0.44$ | $6.44 \pm 0.11$ | $6.15 \pm 0.11$ |
| 7 | 1440 | 1460 | $-3.64 \pm 0.32$ | $-2.90 \pm 0.31$ | $8.77 \pm 0.07$ | $8.22 \pm 0.07$ |
| 8 | 1440 | 1460 | $0.82 \pm 0.32$ | $-0.70 \pm 0.39$ | $8.39 \pm 0.08$ | $8.42 \pm 0.10$ |
| 9 | 1440 | 1460 | $-4.25 \pm 0.54$ | $-5.48 \pm 0.42$ | $7.45 \pm 0.14$ | $9.77 \pm 0.11$ |
| 10 | 1440 | 1460 | $-2.25 \pm 0.38$ | $-4.18 \pm 0.27$ | $6.91 \pm 0.09$ | $7.59 \pm 0.07$ |
| 11 | 1440 | 1460 | $0.37 \pm 0.47$ | $-0.44 \pm 0.26$ | $3.53 \pm 0.12$ | $3.78 \pm 0.06$ |
| 12 | 1440 | 1460 | $-3.21 \pm 0.42$ | $-2.23 \pm 0.35$ | $9.13 \pm 0.11$ | $8.92 \pm 0.09$ |
| 13 | 1440 | 1460 | $0.31 \pm 0.53$ | $-0.84 \pm 0.54$ | $5.28 \pm 0.14$ | $6.37 \pm 0.14$ |
| 14 | 1440 | 1460 | $-2.02 \pm 0.38$ | $-2.28 \pm 0.38$ | $5.67 \pm 0.10$ | $5.53 \pm 0.10$ |
| 15 | 1440 | 1460 | $-1.45 \pm 0.36$ | $-0.10 \pm 0.33$ | $16.10 \pm 0.09$ | $14.95 \pm 0.08$ |
| 16 | 1440 | 1460 | $-16.49 \pm 0.51$ | $-11.70 \pm 0.48$ | $30.24 \pm 0.13$ | $28.94 \pm 0.12$ |
| 17 | 1440 | 1460 | $-2.59 \pm 0.67$ | $-3.88 \pm 0.53$ | $10.74 \pm 0.17$ | $9.61 \pm 0.14$ |
| 18 | 1440 | 1460 | $-2.94 \pm 0.44$ | $-2.61 \pm 0.25$ | $8.68 \pm 0.11$ | $8.54 \pm 0.06$ |
| 19 | 1440 | 1460 | $-4.03 \pm 0.42$ | $-2.55 \pm 0.26$ | $13.24 \pm 0.11$ | $11.87 \pm 0.06$ |
| 20 | 1450 | 1460 | $1.22 \pm 0.49$ | $0.38 \pm 0.30$ | $3.92 \pm 0.11$ | $3.39 \pm 0.07$ |
| 21 | 1440 | 1460 | $-5.17 \pm 0.29$ | $-4.91 \pm 0.41$ | $13.02 \pm 0.07$ | $13.31 \pm 0.10$ |
| 22 | 1440 | 1460 | $-1.21 \pm 0.75$ | $-0.30 \pm 0.28$ | $6.54 \pm 0.18$ | $7.31 \pm 0.07$ |
| 23 | 1500 | 1520 | $-9.83 \pm 0.87$ | $-3.37 \pm 1.02$ | $23.76 \pm 0.20$ | $22.95 \pm 0.23$ |
| 24 | 1455 | 1470 | $-6.26 \pm 0.60$ | $-6.30 \pm 0.60$ | $17.18 \pm 0.14$ | $16.86 \pm 0.14$ |
| 25 | 1440 | 1460 | $-0.81 \pm 0.16$ | $-0.81 \pm 0.15$ | $2.08 \pm 0.04$ | $1.89 \pm 0.03$ |
| 26 | 1440 | 1460 | $1.50 \pm 0.41$ | $1.86 \pm 0.23$ | $5.89 \pm 0.10$ | $6.65 \pm 0.06$ |
| 27 | 1440 | 1460 | $-1.55 \pm 0.36$ | $-3.89 \pm 0.38$ | $9.08 \pm 0.09$ | $10.70 \pm 0.09$ |
| 28 | 1450 | 1460 | $-3.32 \pm 0.42$ | $-3.69 \pm 0.41$ | $11.73 \pm 0.09$ | $11.16 \pm 0.09$ |
| 29 | 1440 | 1460 | $-1.36 \pm 1.27$ | $-1.21 \pm 0.37$ | $6.17 \pm 0.32$ | $6.04 \pm 0.09$ |

Notes.—Quasar rest-frame wavelengths presented in cols. (2) and (3) indicate the spectral ranges used for the blue side of each linear continuum fit. The spectral ranges for the red side are fixed at 1680–1700 Å for each spectrum and are therefore omitted from the table (see § 3.1 for discussion). The slopes (cols. [4] and [5]) and flux density values at $v = 0$ km s$^{-1}$ (cols. [6] and [7]), which are returned by a linear fit for each epoch, are shown. Subscripts 1 and 2 refer to the first and second epochs of observation, respectively.

[a] Object names corresponding to these numbers can be found in Table 1.

rest wavelength at 1549 Å. The velocity translation can thus be summarized with the formula

$$\frac{v}{c} = \left(\frac{R^2 - 1}{R^2 + 1}\right), \qquad (1)$$

where

$$R = \left(\frac{1 + z_{qso}}{1 + z_{abs}}\right), \qquad (2)$$

$z_{abs} = (\lambda_{abs}/1549\text{ Å}) - 1$ denotes the redshift of each part of an absorption feature, and $z_{qso}$ is the SDSS high-S/N redshift of the quasar. The error on the absorber's redshift value is a function of the binning size and is estimated to be $\sigma_{abs} = \Delta\lambda_{abs}/(1549\text{ Å})$. The error in the velocity measurement, $\sigma_v$, was then calculated with standard error propagation of $\sigma_{abs}$ and the error on the calculated quasar redshift, $\sigma_{qso}$, as determined by the SDSS.

A linear least-squares fit in velocity space (in the quasar's rest frame) of the continuum regions chosen for each spectrum was calculated using the POLYFIT routine in IDL. Table 2 also summarizes the fit parameters and their associated error, including the slopes ($m \pm \sigma_m$) and the flux densities of each fit at $v = 0$ km s$^{-1}$ [$f_c(0) \pm \sigma_{f(0)}$]. We recover a linear fit to the continuum,

$f_c(v) = f_c(0) + mv$, with the corresponding flux errors for each velocity bin:

$$\sigma_{f_c(v)} = \sqrt{\sigma_{f(0)}^2 + v^2\sigma_m^2 + m^2 c^2 \sigma_\beta^2}. \qquad (3)$$

### 3.2. Equivalent Width

Normalization provides the best method for removing the effect of continuum variability on the equivalent width measurements of the absorption features. With the spectra translated into velocity space and the continuum fitting complete, each flux-calibrated spectrum is therefore normalized by its respective linear continuum fit. The normalized flux density as a function of velocity in the quasar rest frame is given by

$$f_n(v) = \frac{f(v)}{f_c(v)}, \qquad (4)$$

where $f(v)$ is the SDSS spectrum with the additional flux calibration of Wilhite et al. (2005). The error on the flux density measurement in each velocity bin, $\sigma_{f(v)}$, is simply the respective SDSS error spectrum for each of the quasar observations (Stoughton et al. 2002) after the appropriate scaling as done in Wilhite et al. (2005). These error spectra are also normalized by dividing by

TABLE 3
Two-Epoch BAL Measurements (in Quasar Rest Frame)

| Object[a] (1) | $\Delta t_{qso}$ (days) (2) | $v_{max}$ (km s$^{-1}$) (3) | $v_{min}$ (km s$^{-1}$) (4) | EW$_1$ (km s$^{-1}$) (5) | EW$_2$ (km s$^{-1}$) (6) | $\Delta$EW/$\langle$EW$\rangle$ (km s$^{-1}$) (7) |
|---|---|---|---|---|---|---|
| 1...................................... | 83 | −7600 | −4300 | 2220 ± 261 | 2287 ± 155 | +0.030 ± 0.135 |
| 2a[b]................................. | 105 | −20200 | −10100 | 3570 ± 122 | 4131 ± 175 | +0.146 ± 0.055 |
| 2b.................................... | 105 | −9300 | −3600 | 4536 ± 71 | 4471 ± 97 | −0.014 ± 0.027 |
| 3...................................... | 113 | −10800 | −7400 | 1669 ± 74 | 1544 ± 45 | −0.078 ± 0.054 |
| 4...................................... | 104 | −21300 | −12600 | 3061 ± 91 | 3304 ± 79 | +0.076 ± 0.038 |
| 5[b].................................. | 103 | −9600 | −6200 | 587 ± 95 | 872 ± 60 | +0.391 ± 0.155 |
| 6[b].................................. | 101 | −14300 | −5800 | 2558 ± 125 | 3419 ± 135 | +0.288 ± 0.062 |
| 7[b].................................. | 80 | −14600 | −4200 | 3276 ± 60 | 3814 ± 65 | +0.152 ± 0.025 |
| 8a.................................... | 18 | −17600 | −15900 | 644 ± 34 | 521 ± 41 | −0.211 ± 0.092 |
| 8b.................................... | 18 | −13100 | −5300 | 3790 ± 68 | 4024 ± 85 | +0.060 ± 0.028 |
| 9...................................... | 121 | −11600 | −6300 | 2056 ± 95 | 2017 ± 57 | −0.019 ± 0.055 |
| 10[b]................................ | 118 | −7600 | −5800 | 226 ± 49 | 426 ± 29 | +0.614 ± 0.180 |
| 11[b]................................ | 120 | −12100 | −4600 | 5261 ± 259 | 3740 ± 128 | −0.338 ± 0.064 |
| 12.................................... | 61 | −6900 | −3700 | 695 ± 46 | 848 ± 41 | +0.199 ± 0.080 |
| 13.................................... | 34 | −15100 | −7600 | 4477 ± 226 | 4448 ± 150 | −0.007 ± 0.061 |
| 14[b]................................ | 34 | −16600 | −11100 | 1802 ± 103 | 1760 ± 103 | −0.024 ± 0.082 |
| 15[b]................................ | 36 | −15500 | −11400 | 458 ± 31 | 745 ± 100 | +0.477 ± 0.183 |
| 16.................................... | 19 | −18100 | −3600 | 6890 ± 35 | 6814 ± 34 | −0.011 ± 0.007 |
| 17[b]................................ | 106 | −17100 | −11600 | 168 ± 99 | 1169 ± 83 | +1.499 ± 0.234 |
| 18[b]................................ | 88 | −11600 | −6200 | 1679 ± 76 | 1995 ± 40 | +0.172 ± 0.047 |
| 19[b]................................ | 81 | −14600 | −10600 | 26 ± 41 | 558 ± 27 | +1.824 ± 0.208 |
| 20.................................... | 92 | −15100 | −7100 | 3018 ± 207 | 3481 ± 171 | +0.143 ± 0.083 |
| 21.................................... | 20 | −13600 | −6100 | 1963 ± 34 | 2048 ± 48 | +0.042 ± 0.030 |
| 22.................................... | 79 | −8600 | −6600 | 383 ± 111 | 355 ± 53 | −0.075 ± 0.334 |
| 23a[b]............................... | 31 | −23600 | −17600 | 1746 ± 55 | 1478 ± 70 | −0.166 ± 0.056 |
| 23b.................................. | 31 | −14800 | −10400 | 1715 ± 39 | 1585 ± 50 | −0.079 ± 0.038 |
| 24.................................... | 31 | −21600 | −19100 | 269 ± 34 | 370 ± 35 | +0.316 ± 0.155 |
| 25.................................... | 21 | −9600 | −7200 | 675 ± 67 | 846 ± 72 | +0.225 ± 0.130 |
| 26[b]................................ | 23 | −13600 | −9600 | 794 ± 124 | 1001 ± 53 | +0.230 ± 0.151 |
| 27a[b]............................... | 107 | −17600 | −11600 | 1323 ± 63 | 695 ± 54 | −0.623 ± 0.087 |
| 27b.................................. | 107 | −10600 | −3600 | 4314 ± 60 | 4408 ± 49 | +0.022 ± 0.018 |
| 28[b]................................ | 109 | −17400 | −13100 | 1084 ± 39 | 171 ± 41 | −1.455 ± 0.111 |
| 29[b]................................ | 75 | −10600 | −5000 | 2722 ± 314 | 1737 ± 93 | −0.442 ± 0.147 |

Notes.—Cols. (3) and (4): velocity boundaries of the BAL, shown as vertical dotted lines in Fig. Set 7. Cols. (5), (6), and (7): equivalent widths and fractional change in equivalent width. Subscripts 1 and 2 are used to denote the first- and second-epoch observations, respectively. Object 20 has separate upper and lower error bars, since measurements of fractional change as defined here cannot exceed 2.

[a] Object names corresponding to these numbers can be found in Table 1. For cases in which multiple high-velocity BALs occur in a single quasar spectrum, letters ("a" and "b") are added to the object number to differentiate between the separate features.

[b] Denotes significantly variable BALs, according to the criteria outlined in § 3.3.

continuum flux density (as determined by the fit) to preserve the percentage error in flux density in each bin.

The two-epoch normalized spectra and corresponding normalized SDSS errors are shown in Figure Set 7. The upper and lower rest-frame velocity boundaries of these BALs ($v_{min}$ and $v_{max}$, respectively), shown as dotted vertical lines in Fig. Set 7), between which the flux density does not rise to the level of the fitted continuum, were determined by visual inspection of each spectrum. These BALs and their measured boundaries are presented in Table 3.

Equivalent width measurements of the absorbers were calculated from the normalized spectra as follows:

$$EW = \sum_{i=v_{min}}^{v_{max}} [1 - f_n(i)] \Delta v_i. \quad (5)$$

An additional source of error must be considered when measuring the equivalent widths, since the original flux density has been divided by an approximated continuum fit. The error introduced by fitting propagates into the final equivalent width measurement of each BAL feature,

$$\sigma_{EW} = \sqrt{\sum_{i=v_{min}}^{v_{max}} \left\{ \left[ \frac{\sigma_{f(i)}}{f(i)} \right]^2 + \left[ \frac{\sigma_{f_c(i)}}{f_c(i)} \right]^2 \right\} [f_n(i)]^2 \Delta v_i}, \quad (6)$$

where $\sigma_{f(i)}/f(i)$ is the fractional error of each nonnormalized bin of flux, as determined from the SDSS error spectra, and $\Delta v_i$ is the velocity width of each pixel in kilometers per second.

### 3.3. BAL Variability

To quantify variability in the equivalent widths of the BALs, we measure the fractional change, as given by $\Delta$EW/$\langle$EW$\rangle$, where $\Delta$EW = (EW$_2$ − EW$_1$) and $\langle$EW$\rangle$ is taken to be the average value for the two epochs. Subscripts 1 and 2 refer to the first and second time epoch measurements, respectively. It should be noted that for highly variable BALs these measurements of "fractional change" can appear deceptively small and do not reflect a percentage change. For instance, SDSS J081822.63+434633.8 exhibits a fractional change of 1.824 by this definition, yet its equivalent width has increased by more than a factor of 20.

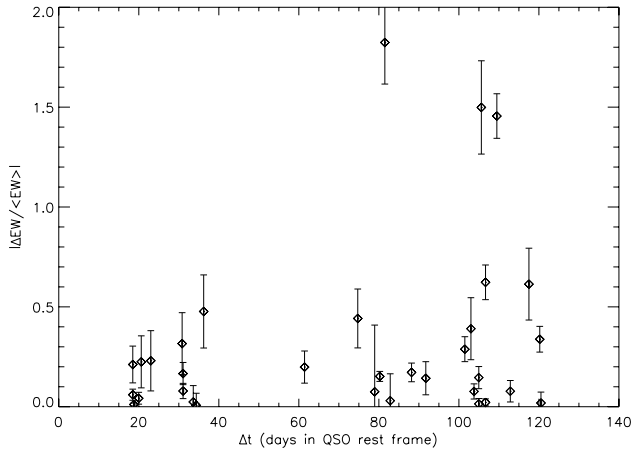

Fig. 1.—Absolute fractional change in BAL equivalent width as a function of epoch separation time, measured in days in the quasar rest frame. Equivalent width variability is strongest for BALs observed over long timescales, indicating the observation of real variability.

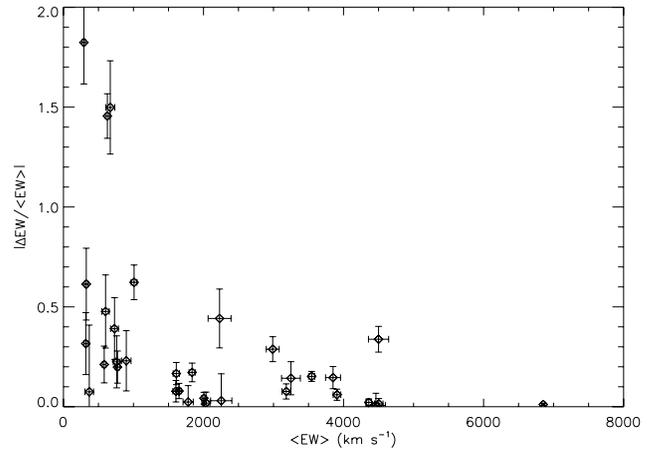

Fig. 2.—Absolute fractional change in BAL equivalent width as a function of the average equivalent width. The greatest variability occurs in the features with relatively small equivalent widths, in accordance with the findings of Barlow (1994).

The absolute equivalent width variability of the 33 BALs in our sample is shown as a function of epoch separation time in the rest frame of the quasar in Figure 1. The positive correlation demonstrates that the most variable BALs have longer epoch separation times, as we should expect when observing real time variability. If the variability distribution had been evenly spread throughout the range of epoch separation times, we might conclude that the observed variability was uncorrelated with time and could be representative of an artifact of the analysis. Instead, we see that the longer the time baseline between observations, the greater the potential for measuring variability, which is a general trend in quasar variability analyses.

Figure 2 shows the absolute fractional change in BAL equivalent width as a function of average equivalent width. We see that the likelihood of detecting equivalent width variability decreases for absorbers with larger equivalent widths, an effect first noted by Barlow (1994). This behavior is what we might intuitively expect, since greater changes in continuum flux are required to ionize absorbers in larger troughs, due to the high optical depths and column densities of strong absorbers. However, continuum flux variation is not required to achieve variability in BALs, as a change in covering factor could produce similar changes in equivalent width and may be necessary to produce variability in the largest saturated features (Arav et al. 2001; Hall et al. 2002). Thus, a change in covering factor could explain the significant variability that we observe in some large equivalent width BALs, while the overall trend favors variability in small equivalent width features.

We also analyze various BAL characteristics as a function of the overall BAL velocity, since absorption features observed at differing velocities are thought to probe geometrically disparate regions within the disk wind (see, e.g., Murray et al. 1995; Proga et al. 2000). We calculate the BAL velocity by taking the difference between the unweighted average velocity of the absorber and the center of the C IV emission line at $v = 0$ km s$^{-1}$ in the quasar rest frame. To investigate any correlation between the location of a BAL within the wind and its tendency to vary, we plot the absolute fractional change in BAL equivalent width as a function of BAL velocity in Figure 3. While a linear relationship between these two parameters is not obvious from these data, we do find that the most highly variable BALs in our sample are found at high velocities, namely, 12,000 km s$^{-1}$ < $v$ < 15,000 km s$^{-1}$.

Having identified equivalent width variability within our sample, we proceed to parameterize the particular ways in which the equivalent widths are prone to variation. Since the changes in BAL equivalent widths can be triggered by bulk motion of the absorber along the line of sight or changes in the optical depth, covering factor, or the degree of ionization in the gas (see, e.g., Barlow et al. 1989; Hamann 1998; Narayanan et al. 2004), we seek to quantify the propensity for variability separately in these parameters. The effects of covering factor, ionization, and optical depth are complexly related, and efforts to separate these effects would require studying the variability of the C IV BAL in relation to broad absorption from other ions in these spectra. Since our study is limited exclusively to the C IV features, the separation of these parameters exceeds the scope of this paper. We instead explore variability in two parameters incorporated in the equivalent width measurement: the velocity width and the average residual flux density, hereafter referred to as the depth. Variability of the velocity width provides information about the bulk motion of the absorber, and monitoring changes in depth summarizes the effects of variation in ionization, covering factor, and optical depth.

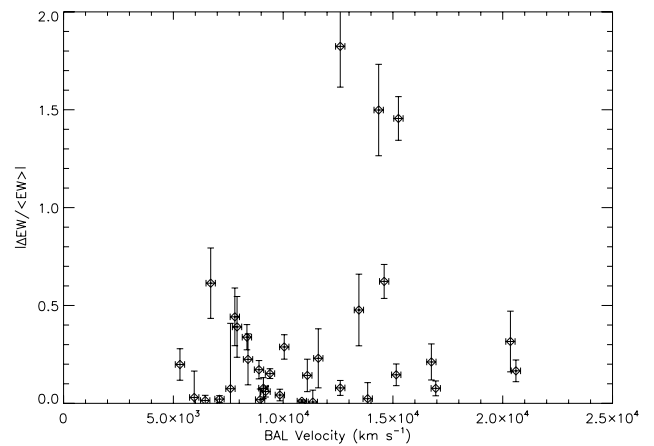

Fig. 3.—Absolute fractional change in BAL equivalent width as a function of BAL velocity. The three most variable features are found at velocities exceeding 12,000 km s$^{-1}$, a region of predicted instability in the Proga et al. (2000) disk wind model.

We define velocity width as the difference between velocity boundaries of the BAL, as determined by a visual inspection and measured in velocity units with respect to the quasar rest-frame emission. These widths have a lower limit of 1000 km s$^{-1}$, as required by our selection criteria, and effectively measure the velocity separation over which the flux density does not rise above the level of the continuum. These boundaries are given in Table 3. Due to the resolution of the spectra in the C IV region for an average redshift of $z \simeq 2$ ($\sim$150 km s$^{-1}$) and the uncertainty in defining the edges of the C IV troughs, we have assigned a conservative error of 300 km s$^{-1}$ for each measurement of the boundary limits in velocity space. Velocity width measurements, therefore, have an associated error of 424 km s$^{-1}$, and errors on measurements of the average BAL velocity are calculated similarly to be 212 km s$^{-1}$.

It should be emphasized that the velocity distribution of our sample is limited by the range of velocities over which C IV BALs may be observed. The Si IV/O IV broad emission line at 1400 Å provides an observational upper limit for determining the C IV BAL velocities, which translates to a velocity of $\sim$28,000 km s$^{-1}$ in the rest frame of the quasar. In addition, our method of continuum fitting further restricts this maximum velocity, since the procedure that we have implemented requires a relatively featureless continuum near 1460 Å, which corresponds to $\sim$19,000 km s$^{-1}$ in the quasar rest frame. The minimum velocity of observed BALs in this sample is further restricted by our inclusion of detached-trough–only features, which exhibit a separation from the emission line of at least 3600 km s$^{-1}$. Thus, we can only select BALs over the velocity interval 3600–19,000 km s$^{-1}$.

The depth of each BAL is recovered by simply dividing the calculated equivalent width of each feature by its measured velocity width at each epoch. This procedure results in a simple unweighted average normalized flux density for each BAL. The uncertainty of the measurement is determined by propagation of the errors assigned to the velocity width and the equivalent width.

As evidenced in Figures 1, 2, and 3, many of the BALs exhibit a fractional change in equivalent width that is consistent with zero. These objects can be omitted from further BAL variability analysis, as they do not aid in characterizing the observed variability. Since one might expect to see variation in all BALs, given sufficiently long timescales, and since most of the BALs in our sample vary to some measurable degree, we can select a "significantly variable" subsample that is large enough to allow for statistical analysis but still limited to include only high-precision data. The criterion for this selection was twofold, allowing significant changes in equivalent width or velocity width to separately qualify a BAL as "significantly variable." BALs exhibiting either a fractional change in equivalent width with at least 2.5 $\sigma$ significance or a fractional change in velocity width with at least 1 $\sigma$ significance were included in this subsample. Fourteen BALs met the first criterion, while seven met the latter. In all, 16 (one-half the total BAL sample) were found to have significant variability by one or both criteria, indicating that significant changes in velocity width are generally, although not always, accompanied by equivalent width variability.

### 3.4. Continuum Variability

Previous studies have reported marginal evidence of a correspondence between changes in the continuum flux and BAL

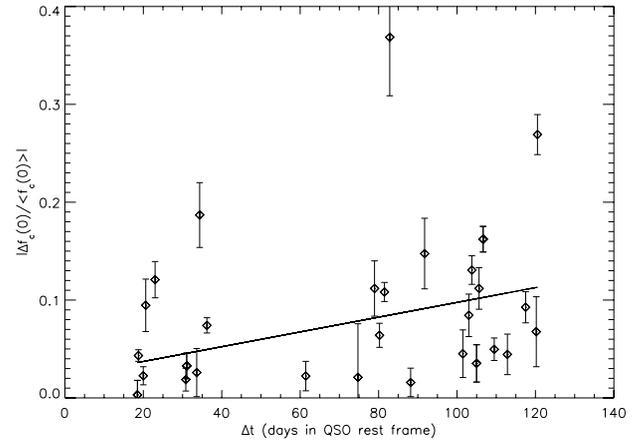

variability (Barlow et al. 1989; Barlow 1994). These results, however, note a high frequency of exceptions to this correlation, which implies that significant variability in a BAL can sometimes be seen in the absence of any change in the continuum flux. Presumably, this disparity results from a time delay between these variations, due to the distance separating the central source and the BAL region, for which lower limits of $3 \times 10^{17}$ to $3 \times 10^{18}$ cm have been speculated (Murray et al. 1995). Given the timescales of our observations, such changes should be frequently out of phase and likely unobservable with only two epochs of observation. Thus, we do not expect to detect coincident variability of BALs with the quasar continuum in our two-epoch data. In addition, changes in covering factor have been suggested as a common mechanism of BAL equivalent width variability, which is independent of variation in the continuum flux (Arav et al. 2001; Hall et al. 2002; Misawa et al. 2005).

To investigate the role of a changing continuum flux in our absorber variability analysis, we have quantified the continuum variability for each object. For our purposes, we define the continuum variability in each object as the fractional change in the flux density of the continuum fit evaluated at 1549 Å ($v = 0$ km s$^{-1}$). This velocity serves as a common reference point approximately midway between fitting regions, which has been chosen for consistency. We quantify the variability of the continuum as the fractional change in fitted continuum flux density at $v = 0$ km s$^{-1}$ as $\Delta f_c(0)/\langle f_c(0) \rangle$, where $\Delta f_c(0) = f_c(0)_2 - f_c(0)_1$ and $\langle f_c(0) \rangle$ is the two-epoch averaged flux density of the fitted continuum, evaluated at the center of rest-frame C IV emission ($v = 0$ km s$^{-1}$). Subscripts indicate the relevant epoch for each measurement.

Using this definition, the variability of the continuum for each object is shown as a function of epoch separation time in the quasar rest frame in Figure 4. A weighted least-squares fit has been overplotted, which shows a convincing increase in variability with increasing time. This trend agrees with previous studies of continuum variability (see, e.g., Vanden Berk et al. 2004; Wilhite et al. 2006) while also revealing the inclusion in our sample of quasars with variable continua. In subsequent analyses, we uniquely identify quasars with significant continuum variability to highlight their differences. These "significantly variable" quasars, 15 in number (one-half the total quasar sample), were identified as such if they exhibited a fractional change

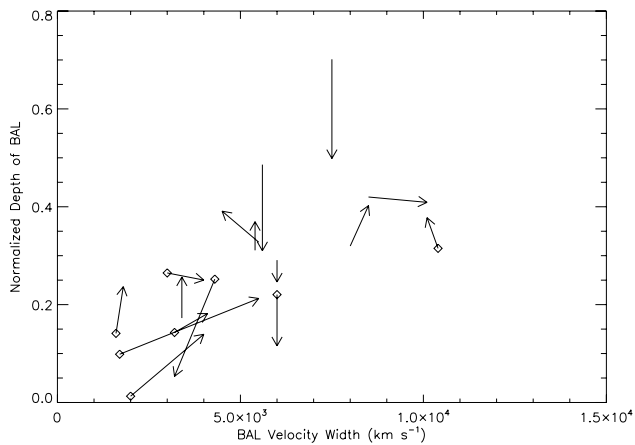

Fig. 5.—Direction and magnitude of changes in depth (in units of normalized flux density) vs. velocity width for the 16 significantly variable BALs. The variable BALs that are associated with the 15 quasars found to have significantly variable underlying continua are marked with diamonds at the first-epoch measurements.

in continuum flux that was more than 4 times the associated error.

## 4. DISCUSSION

Having quantified significant variability among our sample of 33 BALs, we can further explore the ways in which this variability occurs. Figure 5 includes only the 16 significantly variable BALs selected by the aforementioned criteria (see § 3.3). Here we track the changes in the BAL depths and widths with time. Arrows indicate the direction of change with time. In this figure, quasars hosting both significantly variable BALs and significantly variable continua (see § 3.4) are marked at the first-epoch measurement with diamonds. We see that the variable BAL sample draws a nearly equal representation from objects with both variable and nonvariable continua, which gives an indication that our analysis has not been compromised by continuum variability.

The three most variable BALs are associated with variable continua, as we see in Figure 5, but among the larger sample, we do not see a strong correlation between continuum variability and BAL variability. In fact, two of the features exhibiting the greatest change in depth are associated with nonvariable quasar continua. However, the overall propensity of these features to change in depth, rather than in width, suggests that changing levels of ionization in the outflow may indeed be the primary cause of the observed variability, as opposed to a bulk motion of the absorber (Barlow 1994). The argument for ionization-prompted variability is strengthened by the inverse correlation observed between the equivalent width and equivalent width variability among these BALs (as shown in Fig. 2).

If we consider the primary mechanism of variability in these BALs to be changes in the ionizing flux from the central source, we can place limits on the electron densities of each BAL, as shown by Hamann et al. (1997), Narayanan et al. (2004), and Misawa et al. (2005). We do so by taking a nominal gas temperature of 20,000 K (Hamann et al. 1995) and assuming that the BALs are close to ionization equilibrium, with C IV as the dominant ionization state of C. The rest-frame timescales of our observations can then be used to place upper limits on the recombination time for each absorber, and we can extract lower limits of the electron densities (see eq. [1] in Hamann et al. [1997]; eq. [1] in Narayanan et al. [2004]; discussion in Misawa et al. 2005). We thereby calcu-late the limiting electron densities, $n_e \geq 37,900$, $\geq 51,000$, and $\geq 39,000$ cm$^{-3}$, respectively, for the three most variable BALs: SDSS J075010.17+304032.3, SDSS J081822.63+434633.8, and SDSS J093620.52+004649.2.

As seen in Figure 3, the three most highly variable BALs are observed in the velocity range 12,000–16,000 km s$^{-1}$. In Figure 6 we separate the parameters of depth and width to explore these factors of equivalent width variability separately as functions of velocity in the most significantly variable BALs. Observable changes in depth are much more common within the overall sample, as shown in Figure 5, and these changes occur with similar frequency over the entire velocity range of our sample. However, when velocity width variability occurs at a measurable level, it appears isolated within the same high velocity range where the most variable BALs are found. We also notice that the largest changes in velocity width are positive in sign, indicating that the material may be spreading due to changing levels of ionization, covering factor, or acceleration over a range of velocities.

This observation of strong BAL variability at large velocities agrees with the disk wind model of Proga et al. (2000), which predicts the production of variable overdensities in the fast streaming wind as a result of Kelvin-Helmholtz instabilities at ~10,000 km s$^{-1}$. Proga et al. (2000) also predict that such overdensities should reach a maximum velocity of 15,000 km s$^{-1}$ in this stream, which is consistent with our observations. In Figure 5, we see that while many more of the troughs exhibit changes in depth, the most variable BALs all show a change in velocity width as well. This may indicate that, in some BALs, the degree of ionization or covering factor changes as a function of velocity, causing variability in the apparent breadth of the absorbing feature.

It should be emphasized that we have applied a conservative error of 212 km s$^{-1}$ to each BAL velocity measurement ($x$-axis values in Figs. 3, 5, and 6), as explained in § 3.3. These errors are larger than most of the velocity displacements thatt we observe between epochs, which is consistent with the fact that no previous studies of BAL time variability have observed velocity changes greater than 125 ± 63 km s$^{-1}$ on timescales longer than 6 yr in the quasar rest frame (Rupke et al. 2002). As the epoch separations of our data are significantly shorter and our spectral resolution much lower, we should not expect that any apparent changes in velocity in Figures 5 and 6 are significant.

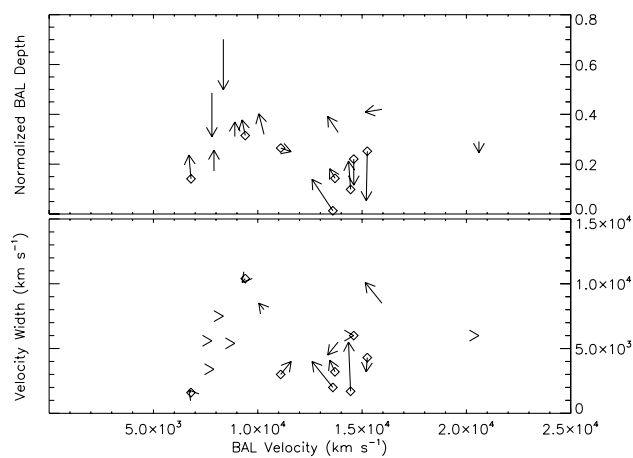

Fig. 6.—Top: Direction and magnitude of change in depth (in units of normalized flux density) vs. velocity for the 16 most significantly variable BALs. Bottom: Direction and magnitude of changes in velocity width vs. velocity for the same BALs. The variable BALs that are associated with the 15 quasars found to have significantly variable underlying continua are marked with diamonds at the first-epoch measurements.

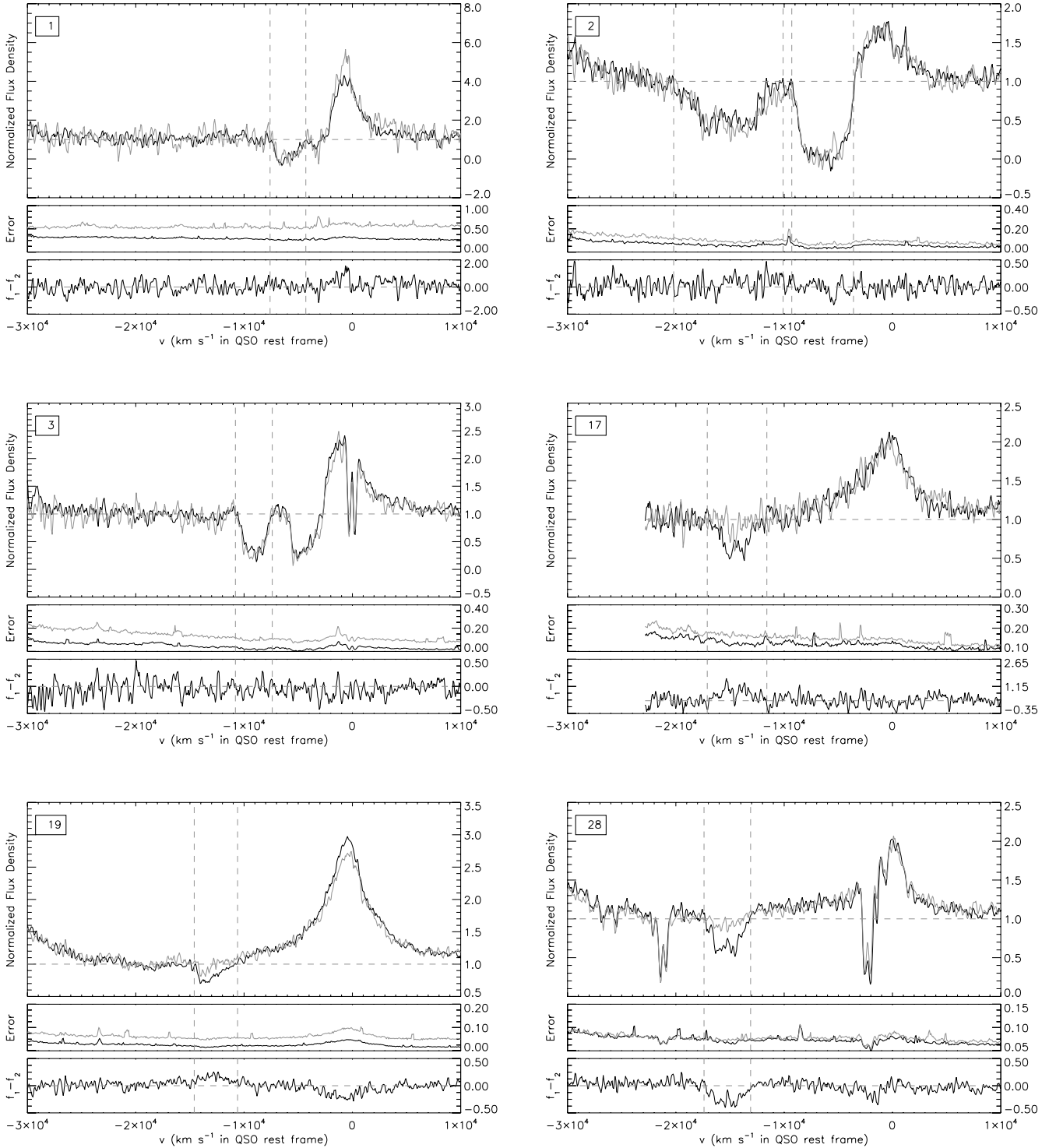

Fig. Set 7.— Continuum-normalized two-epoch SDSS spectral observations for all sources analyzed in this paper. For each source, the top panel presents the high-S/N (*black*) and low-S/N (*gray*) spectra, each with a 3 pixel smoothing. The middle panels show the unsmoothed error spectrum for each epoch, and the bottom panels display the difference spectrum of the two epochs, with a 3 pixel smoothing. Vertical dotted lines overplot boundaries of BALs chosen for analysis. Numbers in the upper left corner of each spectrum correspond to the object numbers used in Tables 1–3. Full object names for each quasar are given in Table 1. [*See the electronic edition of the Journal for Figs.* 7.1–7.29]

## 4.1. *Interesting Objects*

Of the 33 BALs in our sample, three quasars demonstrated exceptionally strong equivalent width variability: SDSS J075010.17+304032.3, SDSS J081822.63+434633.8, and SDSS J093620.52+004649.2. The observed absolute fractional changes in the equivalent widths of these BALs were $1.455 \pm 0.111$, $1.824 \pm 0.208$, and $1.499 \pm 0.234$, respectively. The amplitudes of variability seen in these three absorption features separate them from the other BALs in our sample by a substantial margin, since only one other BAL (SDSS J15026.66−003606.9) was found to exhibit an absolute fractional change greater than 0.70 (see Table 3).

The greatest measured variability within our sample is observed in SDSS J081822.63+434633.8 (object 19 in Fig. Set 7).

The BAL present in this spectrum grows by a factor of 20 in equivalent width between observations. The trough of this BAL is asymmetric, with a noticeably shallower slope on the red side. While the high-velocity end of the BAL deepens between epochs, the velocity width of this feature nearly doubles, as the low-velocity end of the trough extends to absorb more of the continuum flux in the later observation.

SDSS J075010.17+304032.3 (object 28 in Fig. Set 7) hosts the second-most variable BAL, which nearly disappears in the second-epoch observation. The spectrum of this source is especially interesting, as it hosts at least two strong, narrow C IV absorption line systems, presumably also intrinsic to the quasar, that show little if any variability between epochs. The S/N of the second observation nearly matches that of the first; thus, any variability displayed by this BAL is unlikely to be due to noise.

SDSS J093620.52+004649.2 (object 17 in Fig. Set 7) was found to be the third-most variable BAL. In contrast to SDSS J075010.17+304032.3, this BAL appears after being nearly undetectable in the first epoch. This BAL grows symmetrically between epochs, achieving both a greater depth of absorption and a greater velocity width in the later observation.

## 5. CONCLUSIONS

We have detected significant time variability among our sample of 36 detached-trough C IV BALs on time baselines shorter than 1 yr in the quasar rest frame. Included in our sample are three exceptionally variable BALs that are observed in SDSS J075010.1+304032.3, SDSS J081822.63+434633.8, and SDSS J093620.52+004649.2. These three extremely variable BALs are observed within the velocity range 12,000–16,000 km s$^{-1}$. The propensity for variability at these velocities is consistent with recent disk wind modeling by Proga et al. (2000), although other significantly variable BALs in our sample populate a much larger range in velocity (6000–16,000 km s$^{-1}$). Most of the variable BALs in our sample exhibit average equivalent widths of $\leq$2000 km s$^{-1}$, and the three most strongly variable BALs represent some of the smallest features, with equivalent widths of <1000 km s$^{-1}$.

We find that the BALs primarily vary in depth on the timescales that we observe, with fewer than one-half of the significantly variable troughs exhibiting a measurable change in velocity width. This result is consistent with the findings of Barlow (1994) and may indicate that the dominant contribution to the BAL variability is a changing degree of ionization or covering factor. Since we find no strong correlation between changes in BAL equivalent width and continuum variability, it is likely either that the variability of the source is out of phase with the variability of the BAL region or that the changes in covering factor provide the primary mechanism for depth variability in these BALs.

The infrequency of observed changes in velocity likely reflects the very long timescales on which these features can be seen to accelerate. Future higher resolution spectra could improve measurements of velocity variability by enabling us to observe changes within the BAL structure and allowing for direct measurements of the acceleration within the wind. Furthermore, analyzing a similar sample of low-ionization BALs would allow us to explore the variability of BALs at greater outflow velocities, since quasar spectra are receptive to Mg II and Fe II absorption features over a larger portion of the visible spectrum. Future work with more epochs could also place limits on the timescales over which these strongly variable BALs might return to an equilibrium state, and additional monitoring of the three most highly variable BALs may be especially rewarding.

B. F. L., B. C. W., and R. J. B. would like to acknowledge support from Microsoft Research, NASA through grants NAG5-12578 and NAG5-12580, and support through the NSF PACI Project. D. P. S. would like to acknowledge funding from NSF grant AST 06-07634. P. B. H. is supported by NSERC. We would also like to thank the referee for many helpful comments and suggestions.

Funding for the SDSS[8] and SDSS-II has been provided by the Alfred P. Sloan Foundation, the Participating Institutions, the National Science Foundation, the US Department of Energy, the National Aeronautics and Space Administration, the Japanese Monbukagakusho, the Max Planck Society, and the Higher Education Funding Council for England. D. P. S. would like to acknowledge funding from NSF grant AST 06-07634.

The SDSS is managed by the Astrophysical Research Consortium for the Participating Institutions. The Participating Institutions are the American Museum of Natural History, Astrophysical Institute Potsdam, University of Basel, Cambridge University, Case Western Reserve University, University of Chicago, Drexel University, Fermilab, the Institute for Advanced Study, the Japan Participation Group, Johns Hopkins University, the Joint Institute for Nuclear Astrophysics, the Kavli Institute for Particle Astrophysics and Cosmology, the Korean Scientist Group, the Chinese Academy of Sciences (LAMOST), Los Alamos National Laboratory, the Max-Planck-Institute for Astronomy (MPIA), the Max-Planck-Institute for Astrophysics (MPA), New Mexico State University, Ohio State University, University of Pittsburgh, University of Portsmouth, Princeton University, the United States Naval Observatory, and the University of Washington.

[8] The SDSS Web site is at http://www.sdss.org.